# Don't read, just look: Main content extraction from web pages using visual features

GEUNSEONG JUNG, Hanyang University, South Korea
SUNGJAE HAN, KT, South Korea
HANSUNG KIM, Hanyang University, South Korea
KWANGUK KIM, Hanyang University, South Korea
JAEHYUK CHA, Hanyang University, South Korea

Extracting main content from web pages provides primary informative blocks that remove a web page's minor areas like navigation menu, ads, and site templates. The main content extraction has various applications: information retrieval, search engine optimization, and browser reader mode. We assessed the existing four main content extraction methods (Firefox Readability.js, Chrome DOM Distiller, Web2Text, and Boilernet) with the web pages of two English datasets from global websites of 2017 and 2020 and seven non-English datasets by languages of 2020. Its result showed that performance was lower by up to 40% in non-English datasets than in English datasets. Thus, this paper proposes a multilingual main content extraction method using visual features: the elements' positions, size, and distances from three centers. These centers were derived from the browser window, web document, and the first browsing area. We propose this first browsing area, which is the top side of a web document for simulating situations where a user first encountered a web page. Because web page authors placed their main contents in the central area for the web page's usability, we can assume the center of this area is close to the main content. Our grid-centering-expanding (GCE) method suggests the three centroids as hypothetical user foci. Traversing the DOM tree from each of the leaf nodes closest to these centroids, our method inspects which the ancestor node can be the main content candidate. Finally, it extracts the main content by selecting the best among the three main content candidates. Our method performed 14% better than the existing method on average in Longest Common Subsequence $F_1$ score. In particular, it improved performance by up to 25% in the English dataset and 16% in the non-English dataset. Therefore, our method showed the visual and basic HTML features are effective in extracting the main content regardless of the regional and linguistic characteristics of the web page.

## 1 INTRODUCTION

Web pages contain the main content and ancillary areas like ads and navigation menus. Extracting main content is the task of removing peripheries and isolating the primary informative data on a web page. It is used for various purposes, such as information retrieval, search engine optimization, and browsers reader mode [17].

While increasing attention to multilingual and cross-lingual approaches in information retrieval [11, 19], many applications are still unsuitable for non-English texts [1]. In this regard, we would like to know the performance consistency of the main content extraction methods in multilingual situations.

Fig. 1 shows English, Russian, and Japanese web pages and their main contents extracted by two methods: Firefox Readabilitiy.js [13] and Chrome DOM Distiller [7]. While the web pages had a similar subject and layout, the reader modes only worked on the English page. Recent methods like Web2Text [18] and BoilerNet [10] showed their performance only on English web pages.

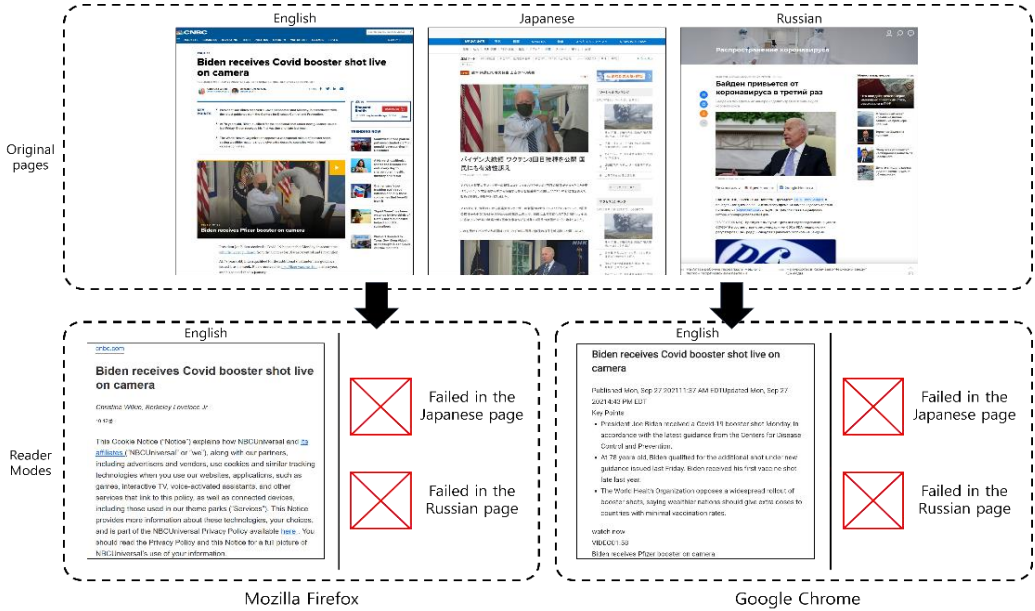

Fig. 1. Browser reader modes' results on pages with similar layout and subject but different languages.

Table 1. Text block matching $F_1$ score in *GoogleTrends-2020* (EN) and *Baidu* (CN) datasets

| Name | EN | Non-EN (Avg.) | Worst |
| --- | --- | --- | --- |
| Readability.js | 0.634 | 0.574 | 0.448 (Japanese) |
| DOM Distiller | 0.399 | 0.441 | 0.294 (French) |
| Boilernet | 0.680 | 0.634 | 0.406 (Japanese) |
| Web2Text | 0.628 | 0.657 | 0.485 (Japanese) |

Table 1 summarizes the extraction performance of the above four methods on the datasets in the ratio of text block matching [18]. These datasets included an English dataset from global websites and seven non-English region datasets in Chinese, Russian, French, Indonesian, Japanese, Saudi Arabian (Arabic), and Korean. The performances decreased in most non-English datasets, and the worst performance was in the non-English datasets. It meant the language and region of the web page affected those methods' performance. Therefore, a multilingual extraction method needed language-independent characteristics rather than the ones used by those existing methods.

For a multilingual extraction method, we used the visual features like the main content's position and appearance. In terms of position, many web pages placed the main content in the middle. However, since the shapes of the main contents are different, it is difficult to determine the exact boundary area of the main content at once. We split this into two steps. First, we found the point close to the main content and then expanded the boundaries for the main content. If we found the point accurately, we can extract the main content with expanding the boundaries from the point. Thus, it is important that selecting an appropriate point and how to expand the boundaries.

There were already two meaningful points: the browser window center and the web document center. We propose a third center for web pages with diverse languages and shapes. This new center is defined as the centroid of the *first browsing area*, which uses the fact that web pages should be considered with user experience. The user's feeling and impression of the web

Don't read, just look: Main content extraction from web pages using visual features page were determined at first sight [20]. When users saw a web page for the first time, their eyes quickly scanned through the page without understanding the texts or content [6, 8]. If they cannot find the web page's main content at once, they may feel uncomfortable with the web page. Since web page authors had to consider making their pages easy to read for users' patterns [15], the main content should be existed or be predictable near the web page initially displayed.

This paper proposed the three steps of the grid–centering–expanding (GCE) method. First, this method created a checkerboard-shaped grid to approximate the first browsing area, called the FBA grid. This FBA grid was at the top of the web page and slightly larger than the browser window to cover the user's instinct web browsing action like mouse scrolling. Second, three center points were determined from the browser window, the web document, and the FBA grid. As these points were expected to close the main content, their nearest text node was highly likely to be a descendant leaf node of the wrapper of the main content. Finally, GCE found a relevant main content area by ascending the DOM tree from the leaf nodes of the previous step. It used the basic properties of HTML elements and visual features, such as the width, height, density, HTML tag name, HTML attributes and distance from the centers. These are not regional and linguistic characteristics, and the values of the properties can be calculated intuitively from HTML and web browsers.

We compared the proposed method with the four existing methods and two Google TabNet DNN models in nine datasets in eight languages (two English datasets in 2017 and 2020). The TabNet models were trained with GCE's visual and basic HTML features. We used Longest Common Subsequence (LCS) and text block matching as performance measures. GCE was 9% to 21% better than the existing methods in LCS $F_1$ score, 16% to 43% better in text block matching $F_1$ score.

In summary, the results of this study make the following contributions:

1. A multilingual method for extracting main content does not depend on web pages' language or regional properties.
2. Helpful features to extract main content for the methods used to traditional rule-based approach and DNN models.

The remainder of this paper is organized as follows. Section 2 presents related works. Section 3 describes the new GCE algorithm proposed here. Section 4 describes constructing new datasets, comparative experiments, and experimental results. Finally, Section 5 presents the conclusions.

## 2  Related Work

Most content extraction methods are based on text and structural-based approaches that have influenced modern browsers' reader mode and recent related research.

Readability.js [13] is a typical rule-based method used by the reader mode of Mozilla Firefox. It examines HTML elements by their tag name, text count, and density of links, along with a text pattern that meets the criteria for main content. It also uses ad hoc rules to check well-known websites' custom tags. It works well on many famous websites.

Boilerpipe [9] is a classifier that uses a decision tree and linear SVM with the shallow text features of continuous text blocks. DOM Distiller [7], the reader mode of Google Chrome, is a hybrid method that uses the Boilerpipe classifier with some rules, similar to Readability.js. DOM Distiller and Boilerpipe focus on the changes in the text features of consecutive text blocks.

Those features include the number of words, text-and-link density, uppercase starting letters, and the structural characteristics of HTML.

Web2Text [18] is a DNN-based classifier that uses a CNN model with features from continuous text blocks. Those features are 128 structural and text-based features, including the word count, the presence of punctuation, and the number of stop words to decide if each text block is part of the main content. However, unlike English, the word counts cannot be calculated by spaces in Chinese and Japanese because they do not use spaces in sentences. Similarly, the punctuation marks like the comma (,) and stop (.) cannot be used for many languages, including Chinese, Japanese, and Arabic, as they have their punctuation marks. The stop words are also entirely language-dependent features.

BoilerNet [10] is also a DNN-based classifier. It uses the LSTM that considers the text node of a web page as a sequence of text blocks whose vectors are words and the path from the root of the DOM tree.

Since many features of Web2Text and BoilerNet are from linguistic characteristics, they may not work correctly in non-English texts. By the way, a visual-based approach can better deal with these linguistic dependencies. The visual-based approach has also been studied for decades. This approach uses features for visual representation like layout, size, and color. VIPS [21] extracts web content structure with visual features. This method segments a web page into *blocks*, the semantic part based on human perception, with 13 heuristic rules about DOM node's structural and style features. It uses text features as visual representation, such as font size and visibility of text nodes, not words or context. For the latest web pages, the rules of VIPS are not available due to changes in the web environment. For example, table nodes like <TABLE>, <TR>, and <TD>, which VIPS treated as basic web page layout components, are no longer used for web page layout. Nevertheless, the visual-based approach is suitable for extracting main content in the non-English web pages because its rules for features are not affected by the web page's language.

## 3  Grid–Centering–Expanding method

The GCE's result was a single HTML element like Readability.js in Firefox. It might include some boilerplate such as ads and navigation blocks. However, the current Adblock methods can filter ads before the page is loaded, and navigation blocks in the main content were usually a "table of content" of the main content, which could be semantic information to humans. Other boilerplates having a "fixed" position in CSS like "Cookie statement" were removed.

To achieve a multilingual main content extraction, we needed to avoid features and methods depending on a specific language. The position was the most important feature in our method. In the sense of human perception, people read top to bottom, and they expect something important to read to appear in the middle. Considering this, we give some definitions (Fig. 2):

Definition 1. **Areas of Browser window and Web document:** The browser window area is the part where the web page is displayed in the browser application on the screen with a window size of $S_w = (w_0, h_0)$. The web document area is the visible part rendered by a browser application with a document size of $S_d = (w_1, h_1)$.

Don't read, just look: Main content extraction from web pages using visual features

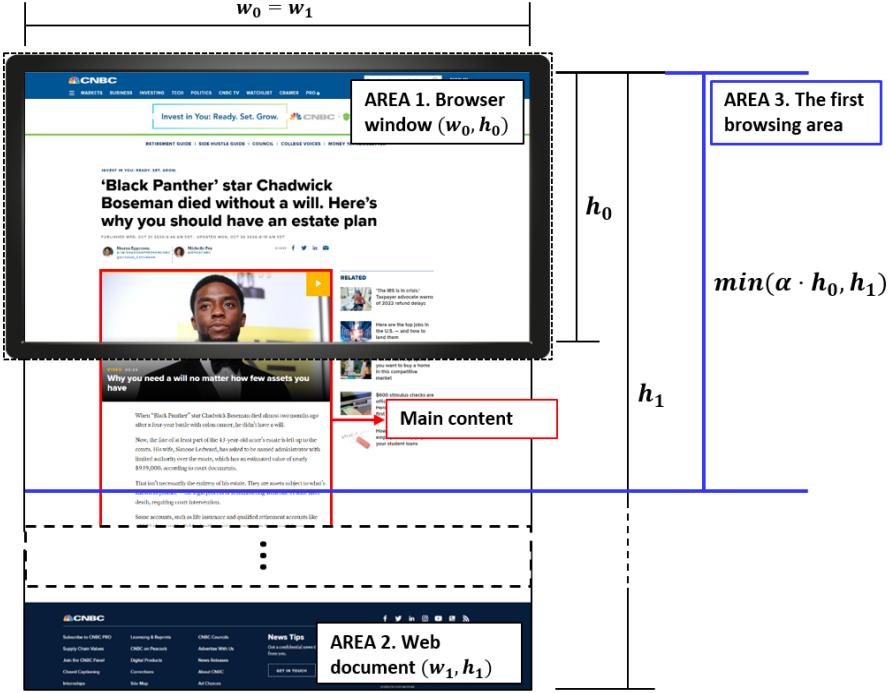

Fig. 2. An example of the first browsing area in a web page. ($8h_0 < h_1$)

Definition 1.1. **The midpoint of the browser window center and the web document center:** The midpoint is defined as the middle point between the browser window center $C_w = (\frac{w_0}{2}, \frac{h_0}{2})$ and the web document center $C_d = (\frac{w_1}{2}, \frac{h_1}{2})$. i.e., the midpoint $C_m$ is $(\frac{w_0+w_1}{4}, \frac{h_0+h_1}{4})$.

Definition 2. **First browsing area:** The *first browsing area* is the visible part of a web document that the user sees first. If $h_0 < h_1$, its size becomes ($w_0$, $\min(\alpha \cdot h_0, h_1)$) with the scrolling threshold $\alpha$ ($\alpha \geq 1$). Otherwise, its size is equal to $S_w$.

The value $\alpha$ is for a user's instinctive browsing behavior like a rough mouse wheel scrolling on the first view.

Definition 3. **Extraction band:** The extraction band is presumed to be relatively more noticeable to the user within the web document. It is the rectangular area between two points, $(0, \min(h_0, h_1))$ and $(0, \max(h_0, h_1))$ as part of the first browsing area.

In our dataset, the midpoint of Definition 1.1 hit the main content in only 80% of web pages, whereas the main content and the extraction band were overlapped in 98% of web pages. We found the more relevant points in the extraction band than the midpoint through the *Grid* and the *Centering* step. In the last step, *Expanding*, we finally extracted the main content by expanding the DOM tree from these points.

In the first step, *Grid*, we devised a grid-style First Browsing Area, the FBA grid. We overlaid a checkerboard-shape grid on the web document. By the definition of the first browsing area, the FBA grid's initial size was $S_w$ and could be extended by adding as many rows as the $\alpha$ value. After the FBA grid size set, some cells that were expected to be irrelevant from the main content were excluded from the calculation centers in the next step.

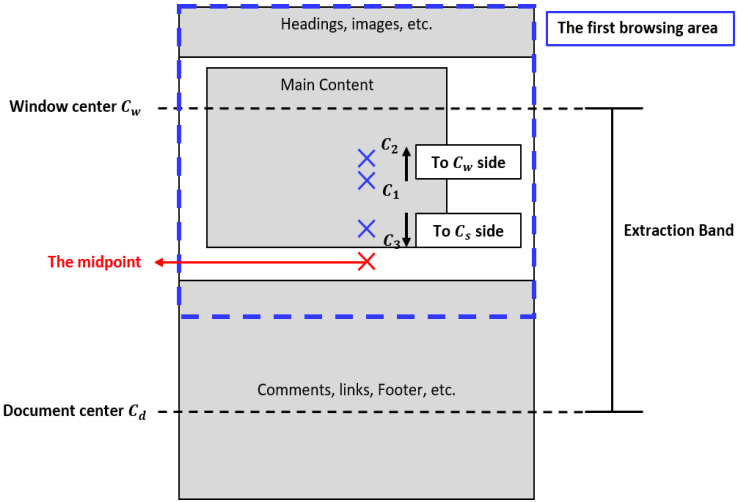

**Fig. 4. An example of *Centering* step when the midpoint cannot hit the main content.**

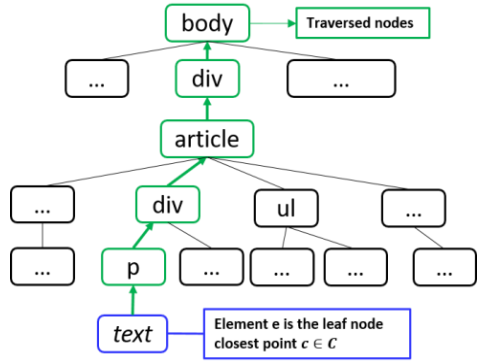

**Fig. 3. A single DOM tree traversal in Expanding Step**

In the second step, *Centering*, we picked points likely to close to the main content with the FBA grid, browser window, and web document area. We determined three center points $C = (C_1, C_2, C_3)$. Since the FBA grid had cells relatively close to the main content, we assumed its center were one of the most relevant points with the main content. $C_1$ was the centroid of the FBA grid. $C_2$ and $C_3$, were the weighted points of $C_1$ with the browser window center $C_w$ and the web document center $C_d$ (Fig. 3). We explain in detail how to compute $C_2$ and $C_3$ in Section 3.2. In most cases, as these points were in the extraction band and the first browsing area, they were highly likely to be in or near the main content than the midpoint. In the next step, we retrieved the main content in the DOM tree with the closest leaf node from each point.

In the third step, *Expanding*, we extracted the main content by traversing the DOM tree with several rules. Because it would be complicated to make a single rule for various languages and shapes, it was more effective to use several rules specialized in each. As shown in Fig. 4, the starting point of the traversal is the closest leaf node from each center point in $C$. Among the first nodes that correspond with each rule, the node with the highest information density became the main content candidate of the traversing round. This traversal was repeated for each center of $C$ so that we could select the final extraction result among three candidates. In section

Don't read, just look: Main content extraction from web pages using visual features

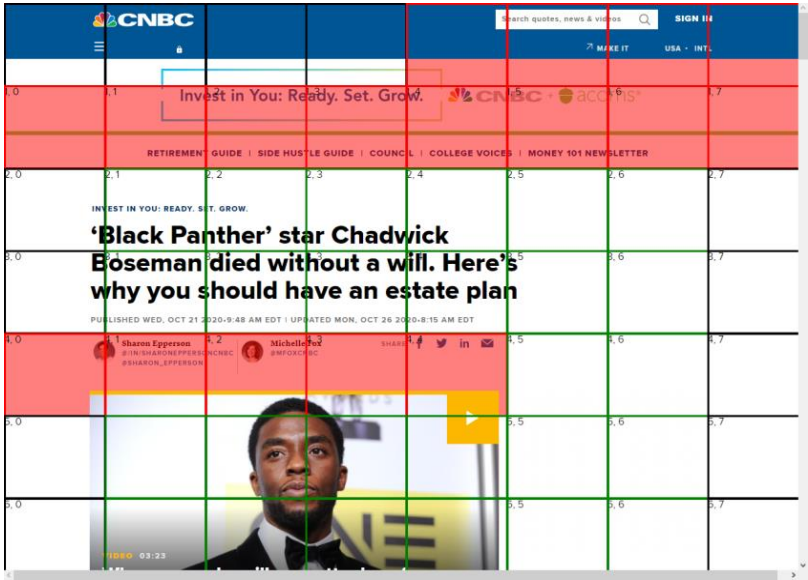

Fig. 5. The FBA grid with excluded cells (red colored cells)

3.3, we explain the three classifiers and the text node area density as an information density criterion.

### 3.1 Step 1: Grid

The FBA grid is a checkboard-shaped grid and consists of rectangular cells. The cell size is determined by dividing the browser window size by the number of rows and columns. For example, the 4 (row) x 6 (column) grid's cell size is 320 x 270 in 1920 x 1080 (Full HD) window. In our experiments, the best values were 7 x 8 grid in 1920 x 1080 display.

As the FBA grid had to cover the first browsing area, its height could expand to ($w_0$, $\min(\alpha h_0, h_1)$). We determined the number of *rows* $N_{row}$ with:

$$N_{row} = ceil\left(\frac{\min(\alpha h_0, h_1)}{h_c}\right),$$

where $h_0$ is the web document height, $h_1$ is the browser window height, and $h_c$ is the cell height. $\alpha$ is a coefficient that simulates a user's unconscious browsing behavior like a fast mouse wheel scrolling looking through a page. This quick behavior makes the growth of the first browsing area downward. The best $\alpha$ in our experiment was 2.

After the FBA grid's size set, we excluded cells expected to be irrelevant from the main content area. The targets were (1) edge cells or (2) overlapped with the navigation menu and ads area. The edge cells were excluded because it was too far from the center and mostly blank (black border cells in Fig. 5). The navigation menu and ads were treated as dense link areas, and many extraction methods filtered them out (red-colored cells in Fig. 5). One of the most famous approach was exploiting <a> tags [9, 17]. The most common link density equations like *textCount* (or *WordCount*) / *linkTextCount* had two problems. First, it underestimated images with links. Seconds, *textCount* and *wordCount* (based on space-separated) values depended on language. For example, the tendency to use a short word in the menu in Chinese, Japanese, and Korean, the *textCount* or *wordCount* approach often missed menu area. In CJK news sites, most words in the top-side menu had two or three letters.

**ALGORITHM 1:** Extract main content candidates $M$

$E \leftarrow$ nodes { $E_1$, $E_2$, $E_3$ }, as DOM leaf nodes
$M \leftarrow \emptyset$
**for** *each $E_n$ in $E$*
    $N \leftarrow E_n$
    $P \leftarrow$ parent node of $N$
    $M_{n\_tag}$, $M_{n\_attr}$, $M_{n\_diff} \leftarrow \emptyset$, Each $M_n$ can be set only once in this loop
    **while** *N is not* `<body>`
        $M_{n\_tag} \leftarrow P$ **if** $P$ **is** `<article>`
        $M_{n\_attr} \leftarrow P$ **if** *attrs(P)* **contains** 'article' **or** 'content'
        $M_{n\_diff} \leftarrow P$ **if** *width(P) > width(N) * r*
        $N \leftarrow P$
        $P \leftarrow$ parent node of $N$
    **end**
    **Add** $M_{n\_tag}$, $M_{n\_attr}$, $M_{n\_diff}$ **to** $M$
**end**
**return** $M$

Instead, we used the link node area density $D_l(e)$ of the visible element e with at least one link descendant.

$$D_l(e) = \frac{A_l(e)}{A(e)}, \qquad (1)$$

where $A(e)$ is the size of e (width * height). and $A_l(e)$ is the sum of the areas of the *link container*s that are descendants of *e*. If a parent node of any `<a>` tag or *link container* element had a single child, the parent became the *link container* recursively. We labeled a text node and its descendants with $D_l(e) > \beta$ into high-density link text node and the rest into low-density link text node. The best $\beta$ was 0.5 in our experiment.

### 3.2 Step 2: Centering

The *Centering* step found the centers likely to lie on the main content with a high probability. According to our observation, the better centers than the midpoint could exist in the extraction band: between the browser window center and the web document center. The centers $C = \{C_1, C_2, C_3\}$ was obtained as follows with the browser window center $C_w$ and the web document center $C_d$.

$$V_a = \{v | v \text{ is a center of a cell in the } FBA\ grid\}$$

$$C_1 = Centroid(V_a), \quad C_2 = Centroid(V_a \cup \{C_w\}), \quad C_3 = Centroid(V_a \cup \{C_w, C_d\})$$

$C_2$ and $C_3$ are the weighted points of $C_1$ to the browser center and the web document center. $C_w$ is for the starting point when users first look at a web page. $C_d$ is for a document with a large height. Although the points $C$ are expected to be close to the main content, not representing the actual HTML element node. Therefore, among the low-density link text nodes, the closest DOM leaf nodes $E = \{E_1, E_2, E_3\}$ are derived from each $C_1$, $C_2$ and $C_3$.

### 3.3 Step 3: Expanding

Ideally, the leaf nodes $E$ from the centers were in the main content. If so, there shuold be a wrapper node of the main content having a node of $E$ as a descendant, and it could be found by expanding the DOM tree from the node of $E$. However, a lack of subtree expansion resulted in a low recall; whereas, excessive expansion resulted in low precision. Therefore, it was necessary to extend the subtree to include the main content as much as possible while excluding irrelevant elements. These rules should not be strict, so there would be a small chance of obtaining no subtree. As we took the first node from the leaf node of $E$ to the root, there was also a small chance of obtaining a large subtree. The rules were applied in the *while* loop of Algorithm 1. We suggested three rules to extract the main content. The rules were (1) the <article> node, (2) the words 'article' and 'content' in the node attributes, and (3) width difference with the parent node. The first two rules have been widely used in other methods and worked well in modern websites. The third rule was from our observation and could be applied to many old or non-English websites.

**<Article> Tag:** $M_{tag}$ `<article>` tag, added in HTML 5, indicates the content is a self-contained composition of the document, page, app, or site. Thus, the content of the `<article>` tag is likely to have readable texts. However, not all web pages are HTML 5, and HTML 5 has not compeled the tag to have the main content. Moreover, it has other uses. For example, if a page had a list of multiple articles like a magazine stand, it was recommended to use `<article>` tag for each article. Nonetheless, this tag is appropriate as the main content feature because using this tag to the main content is the best practice in forum posts, news, and blogs.

**Positive attributes:** $M_{attr}$ The main content area had a specific role in a web page and should be visually separated from other functional areas. Some HTML attributes like *id* and *class* have been commonly used to refer to them. Likewise, web page providers usually assigned some attributes to handle main content areas. This method is effective and widely used. For example, Readability.js increases the weight of the main content possibility of an HTML element if the element has the following positive words: *article, body, content, entry, hentry, h-entry, main, page, pagination, post, text, blog,* and *story.* GCE considers only two words: *article* and *content* because other positive words can imply part of the main content, not whole.

**Width increase:** $M_{diff}$ Traversing the DOM tree from a leaf node to root equals the merging area with sibling nodes. If the main content area had a fixed width, the wrapper of the area should increase downward as the contents (text or image) were merged. However, if horizontal merging occurred, the wrapper had merged with other areas like headers, menus, and footers. Therefore, merging peripheral areas causes a sudden width increase. We stopped the expansion when the width increases over Δw in a single expansion step.

Among the subtree's root selected by each rule, the root with the highest *text node area density* became the candidate $M_n$ of $E_n$. Many other methods had used *text density* as a criterion. For example, the text density of [22] counted the number of text fragments (token) and lines. As we mentioned, the token in some languages cannot be counted the same way as English. Therefore, we counted the area occupied by the text rather than counting the text itself. The *text node area density* $D_t(e)$ of element $e$ is defined as follows:

$$D_t(e) = \frac{\Sigma A(T_e)}{A(e)}, \tag{2}$$

where $T_e$ is a descendant node of *e* with a low-link density and $A(e)$ is the size of e (width * height). $D_t(e)$ cannot be calculated with the $M_{nobest}$ conditions: *M* was <body> node, or the height of *M* was less than half of the browser window height. Otherwise, the $M_{\text{best}}$ with the largest *text node area density* was selected. Consequently, there can be three bests, M1, M2, and M3, from the C1, C2, and C3. We determine the final result in the following order: $M_{3\_best}$, $M_{2\_best}$, $M_{1\_best}$, $M_{3\_nobest}$, $M_{2\_nobest}$, and $M_{1\_nobest}$. This order was better than choosing $M_1$ or $M_2$ first in our experiment. If all *M* are <body> node, the extraction was considered to have failed.

## 4 Experiment

### 4.1 Dataset

Table 2. Experimental datasets.

| Name | URLs | Saved | Readable |
| --- | --- | --- | --- |
| GoogleTrends-2017 | 12720 | 390 | 285 |
| GoogleTrends-2020 | 10560 | 388 | 240 |
| GoogleTrends-2020-KR | 296 | 43 | 21 |
| GoogleTrends-2020-JP | 450 | 47 | 24 |
| GoogleTrends-2020-ID | 900 | 50 | 31 |
| GoogleTrends-2020-FR | 1580 | 97 | 39 |
| GoogleTrends-2020-RU | 890 | 95 | 48 |
| GoogleTrends-2020-SA | 260 | 97 | 43 |
| baidu-2020 (CN) | 1990 | 193 | 53 |

A dataset was built from the GoogleTrends-2017[1] English keywords from global websites, in the same way as Boilernet. We also used the keywords for *GoogleTrends-2020*[2] in the same manner to assess methods with recent web pages. The non-English keywords from South Korea, Japan, France, Indonesia, Russia, and Saudi Arabia are provided by *GoogleTrends-2020-(local)*. The Chinese keywords are from Baidu[3] because *GoogleTrends* has not provided *GoogleTrends-CN*. Each keyword set had 6-12 keyword categories, and the categories varied by year and region. Therefore, the number of collected URLs varied by region, as some regions had few keyword categories. As a result, the top 100 URLs were collected by querying the search engines (Google and Baidu) from the nine keyword sets: two global sets in English and seven local sets in non-English.

Web pages were crawled by designating a certain number of samples from the collected URLs and accessing them with a browser (Table 2). The web page was saved as MHTML with HTML and CSS. As the URL to access was randomly selected, pdf or file links could not be saved as MHTML. From the English datasets, GoogleTrends-2017 and 2020, 400 samples were collected,

---

[1] https://trends.google.com/trends/yis/2017/GLOBAL/
[2] https://trends.google.com/trends/yis/2020/GLOBAL/
[3] https://baijiahao.baidu.com/s?id=1686016936405463174

4Don't read, just look: Main content extraction from web pages using visual features

which was more than that for other datasets, to secure a sufficient training set for TabNet. For all other languages, 50 samples were collected for testing. If there were fewer than 20 readable samples out of 50 samples, another 50 pages were collected.

We labeled ground truth labeling with our browser extension linked to the DOM inspector in the web browser. Our annotators proceeded with labeling according to the following steps: (1) Select any word in the area they thought would be the main content, (2) Using DOM inspector, ascend the DOM tree until the selected node covers all areas of the main content in their mind, (3) Confirm the result and send it to server. If they thought the result was not what they intended in the ascending step, they repeated the steps by selecting another word. We assigned web pages of languages that each annotator could speak.

### 4.2 Evaluation

The GCE's thresholds and values in the experiments were as follows: the browser window size with 1920x1080, 1280x1024, and 2560x1440. The rows and columns of the grid were tried from 3x3 to 8x8 (up to 24x36 grid in 2560x1440). $\alpha$ and $\beta$ were [0.25, 0.5, 0.75] and [1.5, 2, 2.5]. The width increase in the *Expanding* step was tested with [130%, 150%, 170%]. We present the best in the result section with these values: 7x8 grid in 1920x1080 window with $\alpha = 0.5$, $\beta = 2$, and the width increase $\Delta w$ by 170%. With the GCE in the above settings, six methods were used for the experiment. Four existing methods were Readability.js, DOM Distiller, Web2text, and Boilernet. Two Google TabNet models: TabNet<*GoogleTrends-2017-both-80,0.85*> and TabNet<*GoogleTrends-2017-basic-80,0.85*>.

We had also trained DNN models using the GCE's features to verify that the features used were practical for the main content extraction. Google TabNet [3] is a DNN model developed for standardized tabular datasets, which allows instance-wise feature selection during model training through a sequential attention mechanism. We converted web pages into tabular data consisting of columns of visual features and created several binary classifiers using TabNet. We used two feature categories. First, *basic* features were attributes of HTML elements, including *tagname*, *position*, *width*, *height*, *visibility* of the element, *linkCount*, *textCount*, and *descendantsCount*. Second, the *center distance* features were the distances from the browser window center $C_w$ and the web document center $C_d$. We used the global datasets *GoogleTrends-2017* and *GoogleTrends-2020* as the training datasets (a detailed dataset description in Section 5). For example, the TabNet<*GoogleTrends-2017-both-80,0.85*> are created with:

- Dataset: GoogleTrends-2017

- Features (*both*): *basic* features ∪ *Center distance* features

- Ratio of Train (80): Train:Valid:Test = 80:10:10

- Prediction by function `predict_proba`(0.85)

In the same way, the *basic* in TabNet<*GoogleTrends-2017-both-80,0.85*> means the model using *basic* features. We created many classifiers by training dataset, features, and train ratio but presented only two classifiers. TabNet<*GoogleTrends-2017-both-80,0.85*> was the best model, and TabNet<*GoogleTrends-2017-basic-80,0.85*> had the same parameters as the best model with *Center distance* from the *Centering* step. The models were trained with the PyTorch implementation version[4], and the pre-trained models are publicly available.[5]

---

[4] https://github.com/dreamquark-ai/tabnet
[5] https://gitlab.com/hyu_c2s2/main-sweeper

As for performance measures, LCS [17] and text block matching [18] were used. For the LCS metric, $F_1$ scores were calculated as the length of the longest common subsequence by comparing the extracted result with the text of the actual main content. The LCS is intuitive and straightforward. However, it does not tell how close the result was to the actual main content semantically because it does not examine the source of the extracted text node.

Text block matching measure calculates the number of extracted nodes matching the text nodes constituting the main content. We also used $F_1$ scores in this measure. Unlike with LCS, the same texts belonging to different nodes can be distinguished. Nevertheless, the weight between text nodes cannot be calculated. For example, a <p> tag that contains a long text with rich information and a <span> tag of a single word are weighted equally in this measure.

We also assessed the $F_{0.5}$ score of both measures in Appendix. There are two kinds of failures in extracting the main content. If there were no results, the scores would be 0. However, if the result were the entire web page, the recall would become 1, and the F-score would not be 0 even though the extraction failed. We considered the $F_{0.5}$ score to give more penalty to high recall.

### 4.3 Results

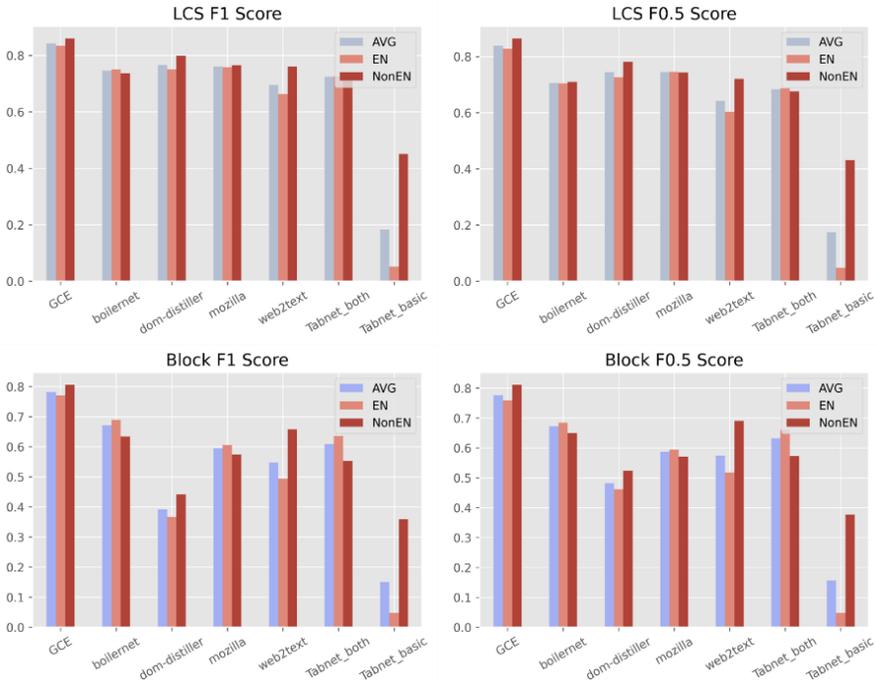

**Fig. 6. $F_1$ and $F_0$ scores by methods**

Fig. 6 shows the evaluation result of LCS and text block matching $F_1$ and $F_{0.5}$. Since the webpages in French, Indonesian, Russian datasets had little difference from the English web pages, the average performance between English and non-English datasets was similar.

In the LCS $F_1$ score in Fig. 6 and Table 3, GCE was better than other methods in all datasets. Most methods had a performance drop in the Japanese dataset. Meanwhile, Web2Text worked better in non-English datasets though it was made for English web pages. It showed high performance in the Chinese dataset. Our Chinese dataset had lacked diversity because most pages were made of Baidu templates. We supposed the high score in the Chinese dataset is why some methods worked so well with the templates. Most methods had a fluctuation in non-English datasets like Boilernet. Thus, it showed that the language of web pages influenced the performance of the extraction method.

Table 3: LCS $F_1$ Score

|  | English (Global) | | Non-English (Local, 2020 year only) | | | | | | | Average | | |
| --- | --- | --- | --- | --- | --- | --- | --- | --- | --- | --- | --- | --- |
|  | 2017 | 2020 | KR | JP | CN | FR | SA | ID | RU | EN | Non-EN | TOTAL |
| GCE | .830 | .838 | .892 | .711 | .906 | .843 | .858 | .870 | .875 | .833 | .859 | .842 |
| Readability.js | .743 | .776 | .784 | .662 | .787 | .729 | .775 | .817 | .773 | .758 | .765 | .760 |
| DOM Distiller | .749 | .751 | .753 | .742 | .876 | .757 | .748 | .837 | .812 | .750 | .798 | .766 |
| Boilernet | .751 | .748 | .887 | .477 | .667 | .755 | .696 | .853 | .828 | .750 | .736 | .745 |
| Web2Text | .651 | .677 | .771 | .714 | .889 | .739 | .690 | .746 | .739 | .663 | .760 | .695 |
| Tabnet_basic | .058 | .042 | .073 | .088 | .818 | .769 | .695 | .086 | .128 | .051 | .451 | .018 |
| Tabnet_both | .724 | .726 | .704 | .535 | .808 | .723 | .687 | .772 | .726 | .725 | .722 | .724 |

Table 4: Text Block matching $F_1$ Score

|  | English (Global) | | Non-English (Local, 2020 year only) | | | | | | | Average | | |
| --- | --- | --- | --- | --- | --- | --- | --- | --- | --- | --- | --- | --- |
|  | 2017 | 2020 | KR | JP | CN | FR | SA | ID | RU | EN | Non-EN | TOTAL |
| GCE | .766 | .783 | .840 | .627 | .860 | .808 | .812 | .769 | .833 | .774 | .805 | .784 |
| Readability.js | .580 | .634 | .511 | .448 | .519 | .534 | .641 | .593 | .687 | .605 | .574 | .595 |
| DOM Distiller | .339 | .399 | .492 | .365 | .653 | .294 | .420 | .316 | .433 | .366 | .441 | .391 |
| Boilernet | .696 | .680 | .834 | .406 | .530 | .703 | .605 | .684 | .716 | .689 | .634 | .671 |
| Web2Text | .382 | .628 | .742 | .485 | .756 | .679 | .597 | .610 | .658 | .494 | .657 | .547 |
| Tabnet_basic | .056 | .037 | .029 | .042 | .702 | .613 | .590 | .048 | .048 | .047 | .359 | .150 |
| Tabnet_both | .646 | .622 | .426 | .357 | .628 | .557 | .611 | .597 | .533 | .635 | .552 | .608 |

The text block matching $F_1$ score in Fig. 6 and Table 4 shows that the reader modes' values were lower than in LCS. Especially, DOM Distiller had the worst performance in the measure because of its text node merging process. For example, if adjacent text nodes existed, such as continuous <li>, DOM Distiller would merge them to create a new single text node. It impaired the score since this evaluation handled the new text node as an unknown node. Though Readability.js uses the traditional rule-based approaches, it was as good as Web2Text using the machine learning approach in overall datasets.

GCE achieved the best performance according to all criteria. Also, it worked well in the Japanese and Saidu Arabian (Arabic) datasets, which were some methods that had a significant decrease in their performance.

Readability.js yielded acceptable performance for the English and non-English datasets. It suggested that high performance was achieved even with traditional rule-based methods through continuous maintenance. However, it had a significant performance drop for the Japanese datasets, revealing a limitation of the rule-based method: an inability to process the characteristics of a specific region.

DOM Distiller was based on a machine learning approach from Boilerpipe and several rule-based approaches similar to the method used for Readability.js. This strategy showed that mixing heuristics and the machine learning approach was beneficial to actual performance.

Boilernet had satisfactory performance, but it shows notably poor performance because of its inability to adapt to the Japanese and the Saidu Arabian (Arabic) datasets. However, with the exceptions of these regions, it showed even scores for all metrics.

Web2Text had relatively inferior performance in general. It was slightly better for the non-English datasets than the English datasets, which suggested that the linguistic features of

Web2Text were not multilingual and not well suited to extracting the main content of web pages. For this reason, it worked well on the Chinese websites with relatively simple tag structures and did not work on the Japanese websites, which had a distinctively developed web environment.

We are not describing the *Center distance* features TabNet models because most were worse than the *basic* models. The TabNet<*GoogleTrends-2017-basic-80,0.85*> had the worst performance, and it was hard to be considered a suitable extraction method. However, the TabNet<*GoogleTrends-2017-both-80,0.85*> had acceptable performance, although the difference between the two models was only *Center distance* features. This result showed that the *basic* features and the *Center distance* features should be used together.

Therefore, the performance of most existing methods had decreased in the Japanese and the Arabic datasets. Although our method had a performance drop in the Japanese dataset, it performed better in all datasets. This result indicated that the proposed GCE method, which focused on visual features, effectively extracted the main web page content.

## 5 Conclusion

On the modern web, though there are web pages in one language (mostly English) accessed by users worldwide, there are also web pages used only by users with specific regions and languages. Nevertheless, the web is similar for all users, and it is undesirable not to be provided with proper technology or services due to regional or linguistic background differences.

The main content extraction methods have been mono-lingual, which works only on English-American web pages rather than multilingual. We have shown that the visual features and the basic HTML features with the simple rule-based approach could extract the main content in an easy-to-understand manner. Since the web pages should be made accessible for people to read, it can be effective to imitate the way people read them. In particular, our method has shown that this approach is useful in a multilingual environment.

## APPENDIX A $F_{0.5}$ SCORES IN EXPERIMENT

Appendix A.1. LCS $F_{0.5}$ Score

|  | English (Global) | | Non-English (Local, 2020 year only) | | | | | | | Average | | |
| --- | --- | --- | --- | --- | --- | --- | --- | --- | --- | --- | --- | --- |
|  | 2017 | 2020 | KR | JP | CN | FR | SA | ID | RU | EN | Non-EN | TOTAL |
| GCE | .821 | .833 | .895 | .708 | .911 | .860 | .854 | .889 | .873 | .826 | .864 | .839 |
| Readability.js | .728 | .765 | .755 | .609 | .758 | .696 | .764 | .814 | .765 | .745 | .743 | .744 |
| DOM Distiller | .722 | .729 | .741 | .712 | .856 | .741 | .721 | .830 | .804 | .725 | .781 | .743 |
| Boilernet | .703 | .705 | .860 | .451 | .654 | .711 | .662 | .842 | .796 | .704 | .709 | .706 |
| Web2Text | .589 | .619 | .705 | .679 | .881 | .676 | .645 | .708 | .676 | .603 | .720 | .641 |
| Tabnet_basic | .054 | .039 | .064 | .003 | .799 | .722 | .661 | .079 | .123 | .047 | .431 | .174 |
| Tabnet_both | .683 | .692 | .627 | .471 | .790 | .687 | .656 | .717 | .653 | .687 | .675 | .683 |

Appendix A.2. Text Block matching $F_{0.5}$ Score

|  | English (Global) | | Non-English (Local, 2020 year only) | | | | | | | Average | | |
| --- | --- | --- | --- | --- | --- | --- | --- | --- | --- | --- | --- | --- |
|  | 2017 | 2020 | KR | JP | CN | FR | SA | ID | RU | EN | Non-EN | TOTAL |
| GCE | .749 | .771 | .840 | .624 | .871 | .832 | .809 | .784 | .825 | .759 | .811 | .776 |
| Readability.js | .567 | .625 | .513 | .435 | .520 | .515 | .632 | .603 | .690 | .594 | .571 | .586 |
| DOM Distiller | .454 | .494 | .571 | .437 | .723 | .400 | .479 | .425 | .521 | .482 | .461 | .482 |
| Boilernet | .690 | .675 | .866 | .406 | .547 | .711 | .599 | .740 | .730 | .683 | .649 | .672 |
| Web2Text | .431 | .620 | .785 | .540 | .803 | .673 | .642 | .650 | .681 | .517 | .691 | .574 |
| Tabnet_basic | .055 | .040 | .027 | .047 | .750 | .644 | .621 | .041 | .042 | .048 | .377 | .156 |
| Tabnet_both | .667 | .652 | .429 | .331 | .697 | .612 | .646 | .578 | .515 | .660 | .573 | .631 |

Don't read, just look: Main content extraction from web pages using visual features